# Intrinsic Optical Properties and Enhanced Plasmonic Response of Epitaxial Silver

*Yanwen Wu[1, 2†], Chengdong Zhang[1†], Yang Zhao[3], Jisun Kim[1], Matt Zhang[1], N. Mohammadi Estakhri[3], Xing-Xiang Liu[3], Greg K. Pribil[4], Andrea Alù[3*], Chih-Kang Shih[1*], Xiaoqin Li[1*]*

1. Department of Physics, University of Texas at Austin, Austin, TX 78712
2. Department of Physics, University of South Carolina, Columbia, SC 29208
3. Department of Electric and Computer Engineering, University of Texas at Austin, Austin, TX 78712
4. J.A. Woollam Co., Inc., 645 M Street, Suite 102, Lincoln, NE 68508, USA

*Plasmonics offers an enticing platform to manipulate light at the subwavelength scale. Currently, loss represents the most serious challenge impeding its progress and broad impact towards practical technology[1-3]. In this regard, silver (Ag) is by far the preferred plasmonic material at optical frequencies, having the lowest loss among all metals in this frequency range[3]. However, large discrepancies exist among widely quoted values of optical loss in Ag[4,5] due to variations in sample preparation procedures that produce uncontrollable grain boundaries and*

[†] These authors made equal contribution

[*] Email: elaineli@physics.utexas.edu, shih@physics.utexas.edu, alu@mail.utexas.edu

*defects associated with additional loss. A natural question arises: what are the intrinsic fundamental optical properties of Ag and its ultimate possibilities in the field of plasmonics? Using atomically-smooth epitaxial Ag films, we extracted new optical constants that reflect significantly reduced loss and measured greatly enhanced propagation distance of surface plasmon polaritons (SPPs) beyond what was previously considered possible. By establishing a new benchmark in the ultimate optical properties of Ag, these results will have a broad impact for metamaterials and plasmonic applications.*

In metals, loss - or absorption - is dominated by intraband (interband) transitions for energies below (above) a certain threshold, which is ~ 3.8 eV in *Ag*. Plasmonic devices are typically operated in the Drude regime below the transition threshold, where loss can be phenomenologically described through the electron scattering rate determined by electron-electron, electron-phonon, and impurity/defect scattering processes. This rate is sensitive to the sample preparation procedure, so loss varies widely for samples of different quality. Loss is concisely described by the imaginary part of the permittivity ($\varepsilon = \varepsilon_1 + i\varepsilon_2$). The two most cited sets of *Ag* optical constants, from papers by Johnson/Christy (JC)[4] and Palik[5], report imaginary parts of permittivity that differ by more than a factor of 3 at visible wavelengths. Measured optical constants for thermally evaporated *Ag* films typically fall in between these two sets of measurements. Curiously, the JC measurements conducted more than four decades ago still represent the lowest loss reported to date for *Ag*, and thus have been widely used to design and model plasmonic devices and metamaterials. With recent rapid advances in plasmonics, it has become of utmost importance to determine the ultimate limitations in the performance of *Ag* as a plasmonic material platform. This can be done by measuring the intrinsic optical constants of *Ag* films of the highest possible quality.

In this work, we performed careful ellipsometry measurements and analyses on atomically smooth, epitaxially grown, single crystalline *Ag* films[6-9], and accurately extracted Kramers-Kronig (K-K) consistent optical constants. Our measurements suggest that the intrinsic loss in *Ag* is significantly lower, by a factor of ~ 2 in the visible wavelength range, than the best values previously reported by JC[4]. Unlike those typically performed on thermally evaporated films, our measurements were not influenced by grain boundary effects. Thus, we were able to fit the experimental data with a simple, three-component analytical model, facilitating the calculation of other important parameters such as the material Q-factor and group velocity. In order to both confirm our retrieved optical constants and to establish the impact of our findings on plasmonics, we measured SPP propagation distances along the *Ag* film, finding extraordinary propagation lengths approaching the fundamental limit determined by the new optical constants at both visible and near-infrared (NIR) frequencies. These results provide a new set of standards for the fundamental optical properties of *Ag*, and have wide implications for metamaterials and other plasmonic applications[10-12]. Simulations of a few key functional plasmonic components using these new optical constants further demonstrate that epitaxial silver may provide a new material platform for significantly improved plasmonic devices, including metasurfaces, surface enhanced Raman scattering, and nanolasers (spasers).

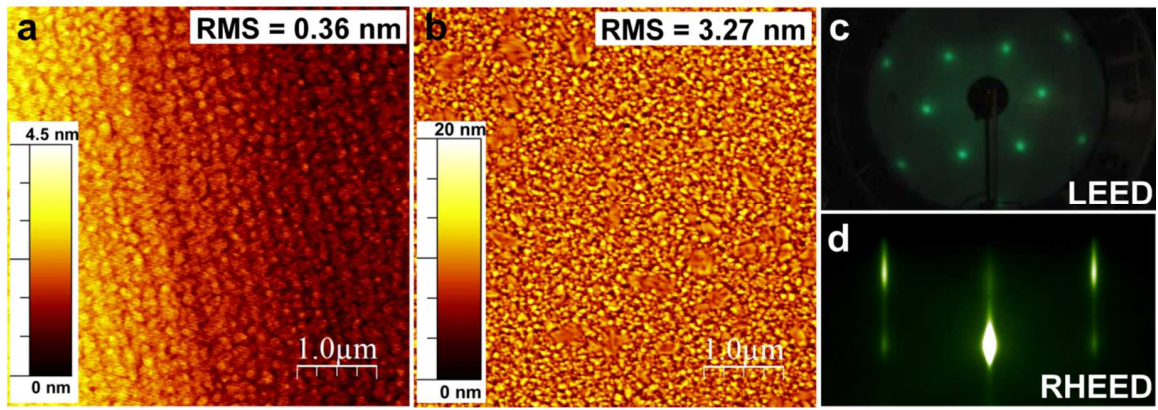

**Figure 1**. AFM scans of (a) a 45nm epitaxial (2 nm $Al_2O_3$/1.5 nm $MgO$ capped) and (b) a 50nm thermally deposited ($0.3$ Å/$s$ deposition rate) silver film. Note the differences in surface height color scales. (c) LEED and (d) RHEED patterns of an epitaxially grown silver film.

The atomically smooth *Ag* films were grown using molecular beam epitaxy (MBE) on heavily doped Si(111)-7×7 substrates. See the method section for details of the growth procedure. The high surface quality of the expitaxial films was confirmed by atomic force microscopy (AFM), shown in Figure 1a, for a 45 nm *Ag* film capped with 1.5 nm *MgO* and 2 nm $Al_2O_3$. For contrast, a 50nm thermally deposited film was also scanned, as shown in Figure 1b. The evaporated film has an RMS roughness of 3.27 $nm$ while the epitaxially grown film has a roughness of only 0.36 $nm$, nearly an order of magnitude smaller. Considering that a single atomic step on an *Ag* (111) surface is 0.24 $nm$, the roughness of these epitaxial films is around the size of 1-2 monolayers. Additional evidence for atomic smoothness is the visibility of sublaminar silicon steps and silver islands even on the surface of a 45 $nm$ thick film, as shown in Fig.1a. Reflection high-energy electron diffraction (RHEED) and low energy electron diffraction (LEED) patterns shown in Fig. 1c and 1d, respectively, taken *in-situ*, further confirm the long-range single crystalline nature of these films.

We used spectroscopic ellipsometry (SE) to extract the optical constants of the epitaxial films. While data collection using ellipsometry is straightforward, the subsequent data analysis is a sophisticated procedure (see Supplementary Information S1). Extracting optical constants from the ellipsometric data is fundamentally an "inverse" problem, where the "results" (polarization ratio of reflected light) are measured and the "cause" (optical constants of the layered structure under investigation) is retrieved to match the measurements. For absorptive thin films, the extracted optical constants and film thickness are inherently correlated[13-15]. Thus, a uniqueness test has to be performed to ensure the integrity of the retrieval process and of its independence from the film thickness (see Supplementary Information S5). For this reason, we chose a parameterized model that adheres to the K-K relations, instead of a tabulated set of independent values as a function of energy. Specifically, the parameterized model used for $\varepsilon_2$ of the *Ag* film as a function of energy E is:

$$\varepsilon_2 = \varepsilon_{Drude} + \varepsilon_{Psemi} + \varepsilon_{Lortz},$$

$$\text{where } \varepsilon_{Drude} = \frac{-A_D * B_D}{E^2 + i\, B_D * E} \text{ and } \varepsilon_{Lortz} = \frac{A_L * B_L * E_L}{E_L^2 - E^2 - i\, B_L * E},$$

The Drude component describes the absorption in the low-energy region, and is essentially a Lorentzian centered at zero energy. $\varepsilon_{Psemi}$ is an empirical term often used to describe absorption near the bandgap in semiconductors[16] and used here to model the interband transition threshold. Its mathematical expression is too complicated to be listed here, though a more detailed description may be found in the Supplementary Material S2. Since interband transitions in metal do not have a well-defined energy gap, an additional Lorentzian term, $\varepsilon_{Lortz}$, centered at $E_L$ is used to provide a more gradual increase in absorption near the *d*-band transition in *Ag*. Obviously, using $\varepsilon_{Psemi}$ and $\varepsilon_{Lortz}$ to model the complicated interband transitions in metal is a simplification. It is, however, a sufficient approximation as evidenced by the excellent fit to the measured data (See Supplementary Information S3 for fit detail).

We performed SE measurements and applied the above analysis to multiple epitaxial and thermal silver films (see Supplementary Information S4 for the complete set of measurements). We present here results from three representative films: an uncapped 40 $nm$ epitaxially grown film, a 45 $nm$ epitaxial film capped with 1.5 $nm$ of $MgO$ and 2 $nm$ of $Al_2O_3$, and an uncapped 50 $nm$ thermal film deposited at a rate of 3.5 $Å/s$ serving as control. The $MgO/Al_2O_3$ cap is a crucial element to prevent the rapid degradation of epitaxial silver due to surface oxidation in ambient conditions, and we confirmed via SE measurements that the $Ag$ film's pristine quality persists even after capping. In our fitting, we used simple *Ag/Si* or *capping/Ag/Si* structural models for the epitaxial films as shown in Figure 2a. There was no silicon oxide layer, as it was removed by an *in-situ* flash-cleaning process. Further silver oxide formation was minimal as all SE measurements were done within one hour of the sample being taken out of vacuum[17]. In either case, adding a thin silver oxide layer to the model did not affect the extracted optical constants of $Ag$. The fitted results for $\varepsilon_2$ of the three films are shown in Figure 2 along with their respective fit residues (see Method for details). Data from each film is plotted against data for the 40 $nm$ film measured by JC and the data compiled by Palik from multiple samples.

We break down the analysis to two energy ranges, as discussed earlier. In the low-energy range (< 3.8 eV), the contribution to $\varepsilon_2$ mainly comes from intraband transitions assisted by scattering between electrons, lattice vibrations (phonons), and imperfections on the surface and inside the bulk of the films. Since epitaxially grown films are single crystalline, scattering from imperfections is expected to be lower than that in the thermal films. This difference is apparent in Figure 2b and 2c where the Drude tails are significantly and consistently lower for epitaxial films than for the control thermal film shown in Figure 2d. More remarkably, our measurements show significantly lower loss than JC's values in the 1.8-2.5eV range, while the control film shows values in between JC and Palik data. The residues in this region are small and centered around zero for all films, suggesting that our model fits the data very well. In the lowest energy range below 1.5 eV, the error becomes larger and our extracted $\varepsilon_2$ appears to be larger than that reported by JC. We note that the original JC paper pointed out large errors in their data in this same energy region (see Supplementary Materials S3 for details). Our extracted optical constants lie well within the range of error, and thus are consistent with the JC results.

In the higher energy range (> 3.8 eV), the extracted $\varepsilon_2$ may not be as accurate. This is primarily due to the approximations of our model, which cannot fully account for the complexity of interband transitions. Including additional Lorentzian functions did not improve the fit, but adversely increased the correlation between fitting parameters[13]. The residues shown in Figure 2b and 2c are not completely random: on close inspection, in the energy range near 3.7eV, indicated by the orange circle in all panels, one can see a sizable peak in the residue for the thermal film (Figure 2d) that is absent in both epitaxial films (Figure 2b and 2c). This peak is associated with the presence of grain boundaries, and we indeed observe the same feature around the same energy level in the JC data. Previous theoretical studies modeling the effect of grain boundaries have suggested that they lead to higher loss in a certain wavelength range determined by their average size[18]. These earlier calculations have, in fact, predicted higher loss in this energy range, consistent with our experimental observation that epitaxially grown films do not show these features.

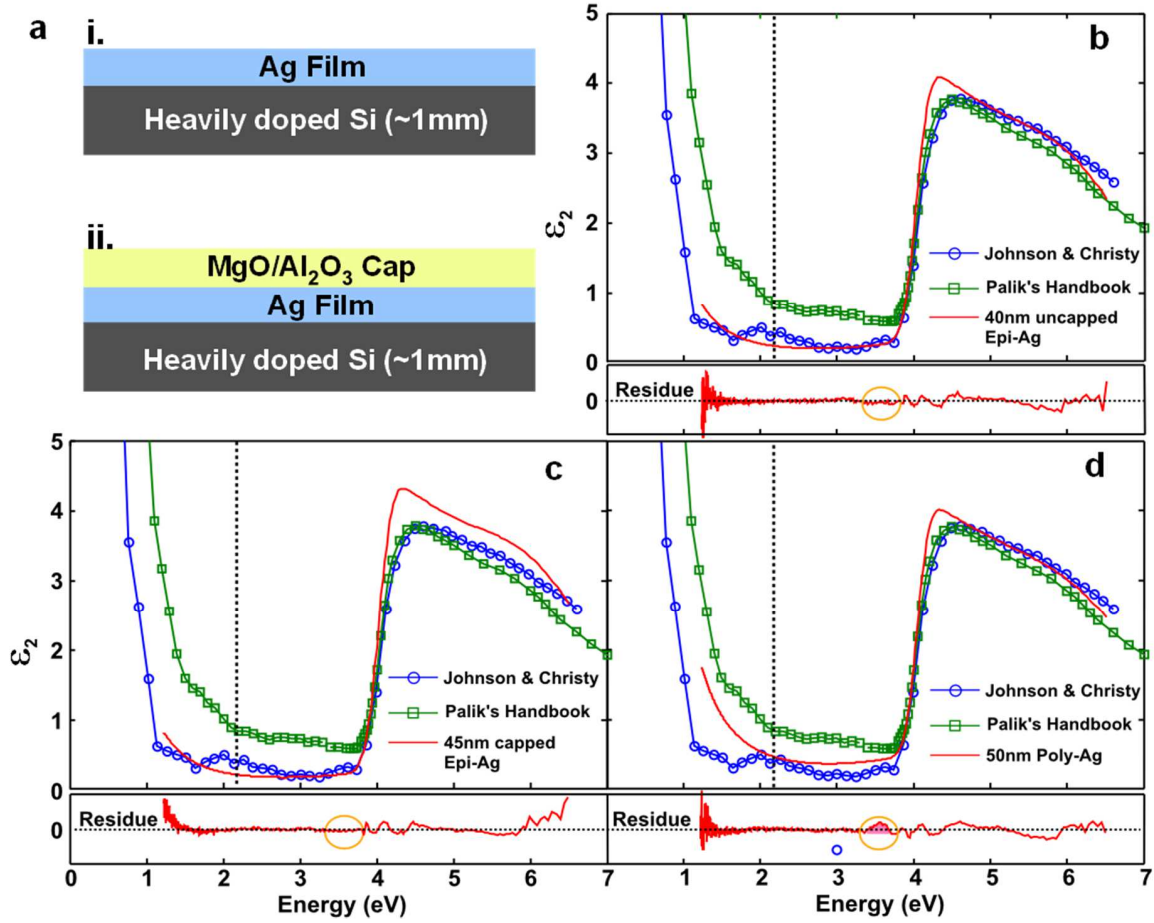

**Figure 2. Spectroscopic ellipsometry (SE) measurements of three Ag films. a.** Layered structures of our *Ag* film samples with and without oxide capping. Energy dependence of $\varepsilon_2$ extracted from SE measurements on **b.** an uncapped 40 nm epitaxial film, **c.** a 45 nm epitaxial film capped with 1.5 nm of *MgO* and 2 nm of *Al$_2$O$_3$*, and **d.** an uncapped 50 nm thermal film deposited at a rate of $3.5 \text{ Å}/s$. Each film is plotted against data taken by Johnson and Christy[4] and from Palik's Handbook of optical constants[5]. The dashed lines in **b-c** indicate the energy at which $\varepsilon_2$ for the epitaxial *Ag* is ~2 times smaller than that of the JC measurements.

Our measured optical constants provide a way to calculate improved theoretical limits to the performance of plasmonic devices. In order to experimentally demonstrate these improvements, we measured extraordinary SPP propagation distances over the 45 nm epitaxial *Ag* film. We excited and detected the SPPs in reflection geometry, as illustrated in Fig. 3a. Light incident from an oblique angle on a single groove launches the SPPs, which are subsequently detected at a series of output coupling slits with increasing distance from the launching site, as shown in the

SEM image in Fig. 3b. We used two different incident wavelengths (632 nm and 880 nm). In both cases, the integrated optical signals from the output grooves are plotted as a function of their separation from the long launching groove as shown in Figure 3c. The experimental data were fitted with simple exponential functions, and we extracted propagation distances of $22 \pm 5\ \mu m$ and $42 \pm 3\ \mu m$ for SPP at 632 nm and 880 nm, respectively. An analytical calculation (see Method for details) using the optical constants measured from the epitaxial film predicts propagation distances of 42 $\mu m$ at 632 nm and 155 $\mu m$ at 880 nm, as shown in the inset of Figure 3c. The remaining difference between our experiment and calculation mainly arises due to scattering from the 1-2 monolayer fluctuations at the surface as shown in Fig. 1a.

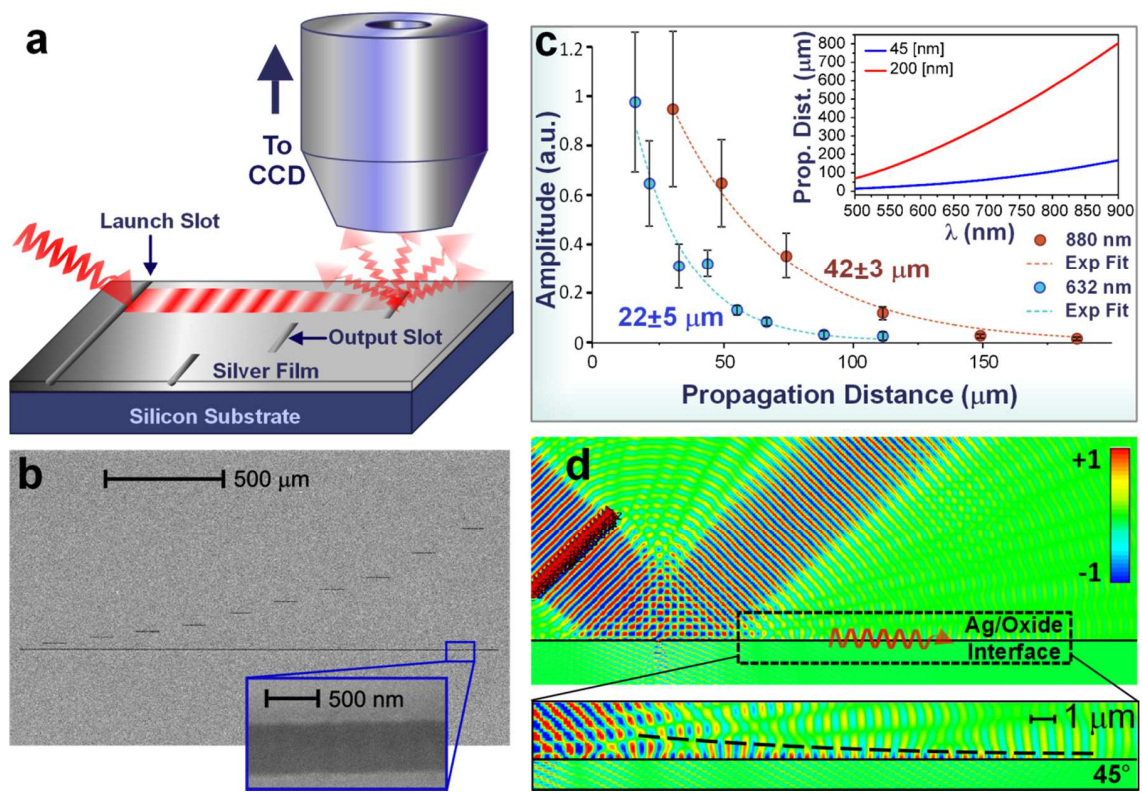

**Figure 3 Schematic, experimental results, and simulations of propagation distance measurements. a.** Schematic of the propagation distance setup. **b.** SEM image of the launching and output slots on the 45 nm thick epitaxial *Ag* film. **c.** Propagation measurements for two excitation wavelengths (632 nm and 880 nm). The data are fitted to exponential curves. The inset compares the simulated propagation distances at two different film

thicknesses using the measured optical constants. **d.** Simulation of the mode profile and the SPP excitation at the interface between Ag and the oxide capping layers.

It is well known that the SPP propagation distance strongly depends on the film thickness. The excited SPP mode in a 45 nm thick film, which is confined at the *Ag*+oxide/air interface, partially resides inside the *Ag* film, but with substantial radiation loss into the substrate due to the thinness of the film. With a film thickness of 200 nm, for instance, the propagation distance would increase to 247 µm at 632 nm and 755 µm at 880 nm, respectively. The SPP launching mechanism and expected modal profile is further verified by conducting a full-wave simulation of the geometry (Fig. 3d), employing the experimentally retrieved optical parameters (see Methods for details). In thermally evaporated films, the presence of grain boundaries limits the SPP propagation distance to a few micrometers in a thick film of 200 nm over the whole visible wavelength range. When a template stripping technique was applied, the SPP propagation distance was shown to improve significantly[19]. The propagation distance extrapolated from our measurements far exceeds the theoretical limit (by a factor of ~ 10) quoted in Ref. 19, which includes an unknown scattering length to account for the effect of grain boundaries.

The experimental realization of inherently lower loss (than previously expected) and the demonstration of extraordinarily long propagation length on epitaxial *Ag* films suggest enormous prospects for epitaxial *Ag* as the ultimate plasmonic platform at visible frequencies. By using the retrieved intrinsic optical constants, we perform theoretical simulations to predict the performance of a few representative structures, including metasurfaces, plasmonic sensors and active plasmonic devices based on these epitaxial layers.

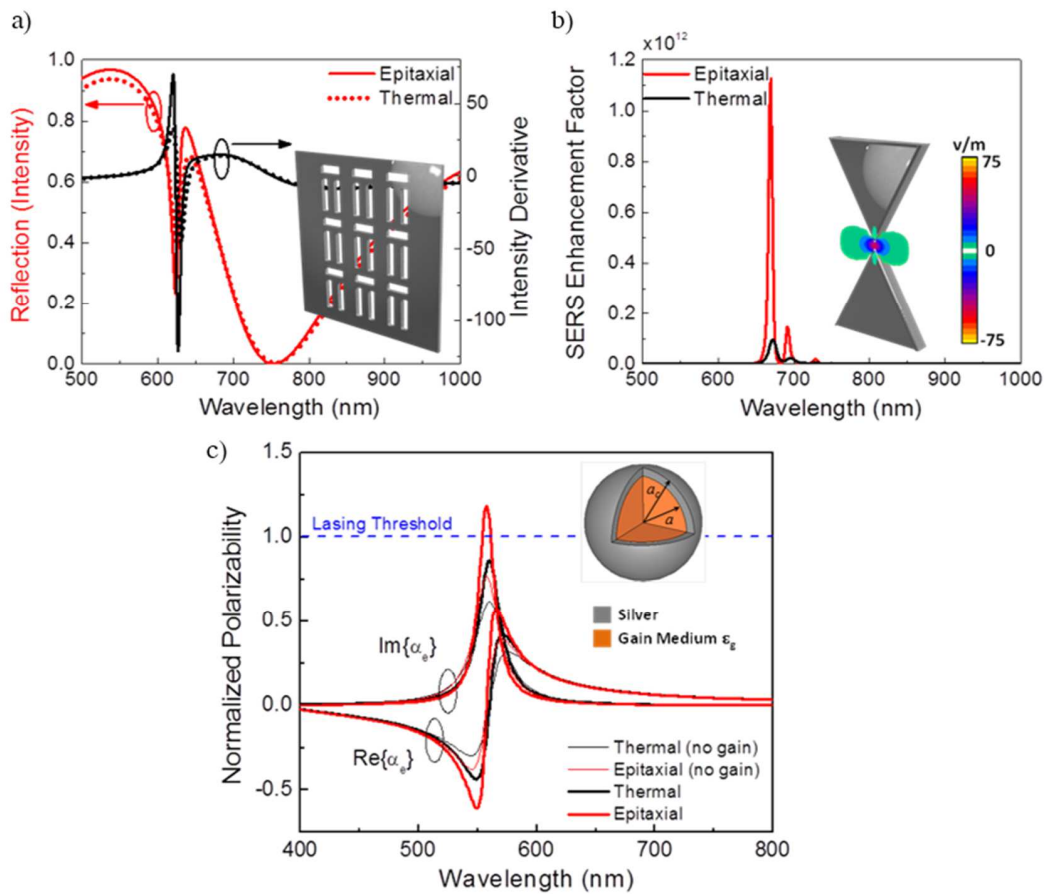

**Figure 4| Designs of resonant structures using epitaxially grown silver film in comparison with thermally grown silver. a.** Reflectance of a dolmen Fano metasurface (red, left y-axis), and its derivative (black, right y-axis) with respect to wavelength. The derivative displays a much steeper slope on the metasurface made from epitaxially grown silver. The unit cell dimension is 180 nm x 191 nm, the dipole antenna has a width of 30 nm and a length of 150 nm, the double antennas have a length of 101 nm, the gap between the double antenna and the dipole antenna is 20 nm, and the gap in the double antenna is also 30 nm. **b.** SERS enhancement factor measured at the gap of a bowtie antenna. The bowtie antenna is composed of two equilateral triangles with arm lengths of 40 nm and a gap of 1 nm. The inset shows the geometry and the simulated electric fields. **c.** Calculated normalized polarizability for a core-shell nanoparticle formed by a core dielectric (with and without embedded gain) and a thin silver shell, comparing epitaxial and thermally grown films. The blue dashed line indicates the lasing threshold.

Our first example (Figure 4a) shows a plasmonic metasurface composed of subwavelength dolmen-like nano apertures[20] with dimensions outlined in the caption of Figure 4. This particular metasurface supports a high-Q spectral feature known as Fano resonance[21,22], a result of the interference between the bright dipole and the dark quadrupolar modes. The Fano resonance appears around 620 nm, clearly visible in the calculated reflection spectra (red curves, Figure 4a left y-axis). The use of epitaxial films only alters spectral properties near the Fano response, where fields are more concentrated in the metal and thus benefit more strongly from reduced losses. The Fano features are significantly sharper in the epitaxial case, as confirmed by calculating the derivative of the reflection intensity with respect to wavelength (black curves, Figure 4a right y-axis). We find significantly larger slopes near the Fano resonance of the metasurface with an increase of about three times in the derivative, a quantity of great interest for sensing applications.

Figure 4b shows another example, this time involving surface-enhanced Raman scattering (SERS)[23], where the enhancement factor can be estimated as $\rho(\omega) = \left| \frac{\mathbf{E}_{loc}(\omega)}{E_{inc}(\omega)} \right|^4$ [24]. In this case, we focus on a bowtie nanoantenna formed by two equilateral triangles with a molecule placed in the middle of the 1 nm wide gap. The difference in material losses, which is approximately a factor of two, translates into a $\sim 2^4 = 16$ times difference in SERS enhancement factor at resonance, as expected.

Finally, Figure 4c shows the application of our epitaxial silver film to spasers[25-27]. For simplicity we analyze a canonical spaser, in the form of a core-shell spherical nanoparticle formed by a semiconductor gain medium with inverted Lorentzian dispersion $\varepsilon_g = \varepsilon_\infty + \frac{\omega_p^2}{\omega^2 - i\omega\gamma - \omega_0^2}$, where $\varepsilon_\infty = 2.0, \omega_0 = 2\pi\, 550\,\text{THz}, \gamma = 2\pi\, 80\,\text{THz}$. The gain medium is covered by a thin $Ag$ shell, with geometry $a_c = 40$ nm, $a/a_c = 0.85$. We plot the frequency dispersion of the normalized polarizability $\alpha / \left[ 6\pi\varepsilon_0 / k_0^3 \right]$, where $k_0$ is the free-space wavenumber and $\varepsilon_0$ is the corresponding permittivity. The polarizability essentially measures the scattering efficiency of the particle around its resonance frequency. We tuned the geometry to support a plasmonic resonance around the material resonance $\omega_0$ of the gain material. The thinner lines in the figure show the case with no gain ($\omega_p = 0$) for reference, and the thicker lines refer to the case when a moderate gain factor is considered ($\omega_p = 2\pi\, 100\,\text{THz}$). It is seen that the epitaxial silver in both cases is able to boost the resonance lineshape and increase the overall polarizability around the plasmonic resonance. Even more interestingly, when gain is considered in this geometry, the core is able to both compensate for the epitaxial silver losses and to boost the level of normalized $\text{Im}[\alpha]$ above the lasing threshold $6\pi\varepsilon_0 / k_0^3$ at resonance, which is a fundamental limit for passive nanostructures. In contrast, lasing cannot be achieved in the proposed core-shell structure when higher loss in *Ag* is incorporated. Our calculation suggests that epitaxial silver may be ideal in realizing efficient subwavelength lasers based on plasmon resonances, due to the reduced level of losses and absorption, and therefore the reduced level of gain required to lase. The qualitative conclusion from our simple model calculation is consistent with the recently demonstrated[28] superior performance of a spaser based on epitaxial *Ag* films.

We highlight two caveats in interpreting the results of the above simulations. First, we evaluated the improved performance of these devices by simply replacing the optical constants reported by JC with the new optical constants extracted from our epitaxial films. In practice, most thermally evaporated films exhibit loss significantly higher than that reported by JC in the optical frequency range. Second, the new optical constants do not capture all advantages offered by epitaxially grown silver films. In the case of SPP propagation distance, the elimination of grain boundaries increases propagation distance by two orders of magnitude compared to that found in thermal films. This increase in SPP propagation distance is partially responsible for the reduced mode volume and remarkably low lasing threshold of plasmonic nanolasers based on epitaxial $Ag$ films[28,29]. These two factors suggest that the overall advantages of using epitaxially grown silver in plasmonic applications are more substantial than what is captured in the above calculations.

We suggest that future theoretical calculations on metamaterials and plasmonic devices based on $Ag$ should incorporate the new optical constants reported here (see Table S2 in Supplementary Information S6 for a list of numerical values), as they better capture the fundamental properties of high quality, atomically smooth silver films and represent the ultimate limitation of bulk $Ag$ for plasmonic applications. While the reported optical constants would not apply to low-quality films produced using thermal evaporations, $Ag$ nanoparticles synthesized using wet chemical procedures are considered to be of crystalline structure, for which these new optical constant values are expected to apply. We anticipate that these high-quality epitaxial silver films and their improved fundamental optical properties will have a significant positive impact on the fields of plasmonics and metamaterials.

**Methods**:

**Sample growth:** All films were grown using 6N (99.9999%) pure silver. To ensure crystalline growth, the silver films were prepared using a refined two-step growth process. First, a particular amount of silver (usually 20 ML) was evaporated onto a liquid nitrogen cooled $Si$(111) substrate (~90 K) with a low deposition rate of ~1 Å/$min$. Then, the sample was naturally annealed to room temperature. The number of iterations of this two-step process determined the final desired film thickness. A commercial Knudsen cell was used as the silver evaporator to ensure a precise and stable deposition rate. The thickness of the films was determined through a calibrated deposition rate measured by a quartz crystal monitor. In order to prevent the formation of an oxide layer and to suppress de-wetting in an ambient environment, we also developed a procedure for *in-situ* layer growth of a transparent capping material *MgO* followed by another 2 nm *Al$_2$O$_3$* capping layer deposited in a separate atomic layer deposition (ALD) chamber. Note that this capping method differs from our previously reported capping method using *Ge* layers[28] (a low bandgap material) which causes additional absorption in the visible wavelength range.

**SE measurements** were conducted with a JA Woollam M-2000 ellipsometer, with focusing probe attachments providing an incident spot size of 300 $\mu m$. Modeling and analysis were performed with the ellipsometer WVASE32 software. We specifically measured the optical constants of the flash-cleaned silicon substrate used in the deposition process to eliminate any discrepancy and uncertainty introduced by the substrate. These measured silicon optical constants are fixed in the subsequent data fitting for all samples, and only the silver parameters are allowed to change. Each film was measured under 3 different incident and collection angles (55°, 70° and 75°) with respect to the sample surface normal. For each angle, data were collected at 6 different film locations and showed no location dependence in the raw data, meaning the silver film is spatially homogeneous. In order to verify that the silver is optically isotropic, we extracted the effective optical constants directly from the raw data before the process of reverse fitting (see Supplementary material for detail). In this case, we saw no angular dependence in the effective optical constants, which confirmed optical isotropy and suggested that multi-angle analysis would not contribute further information on these films. Therefore, we can safely perform the reverse fitting process on data taken at 70° without loss of generality. The residues plotted in Figure 2b-c are the differences ($\varepsilon_{2eff}^{exp} - \varepsilon_{2eff}^{model}$) between the calculated effective optical constant, $\varepsilon_{2eff}^{model}$, and the experimental effective optical constant, $\varepsilon_{2eff}^{exp}$.

After fitting the experimental values of absorption to our model for $\varepsilon_2$, the values of $\varepsilon_1$ are unambiguously determined using the K-K relations (results shown in Supplementary Information S3). When calculating $\varepsilon_1$ from our finite range of collected data, we modeled the properties outside of this range in the form of two additional effective poles, one at lower energy and the other at higher energy.

**Propagation distance measurement**: In order to excite the SPP mode in reflection geometry, a laser beam with spot size of 10 $\mu m$ is focused to various spots along a long launching groove milled on the surface of the silver using a focused ion beam (FIB) as shown in Figure 3a and 3b. The beam is linearly polarized with equal amount of s- and p-components. The shorter output grooves are milled at varying distances (from 5 $\mu m$ up to 200 $\mu m$) from the launching groove. A 50X microscope objective, moving independently from the sample, collects the light from the output grooves due to the decoupled SPPs. The collected light passes through a polarizer that is set to be crossed-polarized to the excitation laser to minimize scattered light and is then imaged on the CCD.

**Calculation of the SPP mode**: The primary characteristics of the SPP mode is studied analytically by exactly solving the mode profile of the structure shown in Figure 2a (ii). The epitaxial film can be viewed as a plasmonic waveguide formed by a silver core, sandwiched between asymmetric dielectric layers of oxide and silicon. The wave equations are then exactly solved inside this waveguide in order to find the dispersion relation based on the experimentally measured optical constants. In the full wave simulation presented in Figure 3d, the SPP energy is shown to strongly concentrate on the metal-air interface and exhibits an exponential decay. We note that the field distribution is modulated by the input laser beam, especially in the proximity of the source spot. Such modulation is also observed in our experimental data, which exhibit small oscillations superimposed on an exponential decay. We used a simple exponential function to extract the propagation distance from experimental data. In addition, we chose to show the field distribution at 45° incident angle for clarity of display. The simulated results for a 63°of incident angle are qualitatively the same.

ACKNOWLEDGMENT

We gratefully acknowledge financial support from the following sources: NSF DMR-0747822, DMR-1306878, Welch Foundation F-1662 and F-1672, AFOSR FA9550-10-1-0022, ARO-W911NF-08-1-0348, ONR-N00014-10-1-0942.

# Online Supplementary Materials for "Optical Constants of Atomically Smooth Epitaxial Silver films and Their Potential for Plasmonic Applications"

*Yanwen Wu[1,2†], Chengdong Zhang[1†], Yang Zhao[3], Jisun Kim[1], Matt Zhang[1], Xing-Xiang Liu[3], Greg K. Pribil[4], Andrea Alù[3*], Ken Shih[1*], Xiaoqin Li[1*]*

1. Department of Physics, University of Texas at Austin, Austin, TX 78712
2. Department of Physics, University of South Carolina, Columbia, SC 29208
3. Department of Electric and Computer Engineering, University of Texas at Austin, Austin, TX 78712
4. J.A. Woollam Co., Inc., 645 M Street, Suite 102, Lincoln, NE 68508, USA

**S1: Basic principles of ellipsometry measurements**

Commercially available ellipsometers are capable of acquiring a high density of precise data. Coupled with flexible and sophisticated analysis software, it is a powerful tool for accurately determining the optical constants of materials. In general, the ellipsometer measures the complex reflectance ratio between *s*- and *p*-polarized light, $\rho = \frac{r_p}{r_s} = \tan\psi\, e^{-i\Delta}$. The raw data output of the ellipsometer is in the form of $\psi$ and $\Delta$. In order to extract the complex optical constant, $\varepsilon = \varepsilon_1 + i\varepsilon_2$, we need a set of initial values of optical constants for each material and a model of the entire layered structure. Starting from these initial values and models, the software calculates $\psi$ and $\Delta$ using the Fresnel equations, then compares them directly to the measured values to yield a mean square error (MSE). Next, the fitting parameters are adjusted, and the analysis process is reiterated until the MSE or the change in MSE is below the preset values. The integrity of the fit depends on the initial model/values of the optical constants. Section S2 describes the particulars of our chosen model, and in section S5 we investigate its reliability.

---

[†] These authors made equal contribution

[*] Email: elaineli@physics.utexas.edu, shih@physics.utexas.edu, alu@mail.utexas.edu

**S2: Ellipsometric model**

In any real material, $\varepsilon_1$ and $\varepsilon_2$ have a causal relationship specified by the Kramers-Kronig (K-K) equations. This means that we can focus on modeling only $\varepsilon_2$, the absorptive component of the dielectric function, and use that to calculate what $\varepsilon_1$ would have to be. The parameterized model used for $\varepsilon_2$ of the Ag film as a function of energy E (in eV) is:

$$\varepsilon_2 = \varepsilon_{Drude} + \varepsilon_{Psemi} + \varepsilon_{Lortz},$$

$$\text{where } \varepsilon_{Drude} = \frac{-A_D * B_D}{E^2 + i\, B_D * E} \text{ and } \varepsilon_{Lortz} = \frac{A_L * B_L * E_L}{E_L^2 - E^2 - i\, B_L * E},$$

The Drude component describes the absorption in the low-energy region, and is essentially a Lorentzian centered at zero energy. $\varepsilon_{Psemi}$ is an empirical term often used to describe absorption near the bandgap in semiconductors, and will be described shortly[17]. Since interband transitions in metal do not have a well-defined energy gap, the additional Lorentzian term, $\varepsilon_{Lortz}$, centered at $E_L$ is used to provide a more gradual increase in absorption near the *d*-band transition in Ag. Obviously, using $\varepsilon_{Psemi}$ and $\varepsilon_{Lortz}$ to model the complicated interband transitions in metal is a simplification. It is, however, a sufficient approximation as evidenced by the excellent fit to the measured data. After fitting the experimental values of absorption to our model for $\varepsilon_2$, the values of $\varepsilon_1$ are unambiguously determined using the K-K relations (results shown in Supplementary Information S3). When calculating $\varepsilon_1$ from our finite range of collected data, we modeled phenomenon outside of this range in the form of two additional effective poles, one at lower energy and the other at higher energy.

The $\varepsilon_{Psemi}$ term in the model of $\varepsilon_2$ presented in the main paper is an asymmetric function that consists of four spline polynomials - two for the right side and two for the left side - used to model the abrupt change in absorption around the energy region of the band edge in semiconductors. While the concept of band gap does not apply to metals such as silver, the Psemi model is still valid in modeling the sharp interband transition threshold. The exact expression of the $\varepsilon_{Psemi}$ term is proprietary. We describe the physical meaning of the seven variables involved and quoted below: $A_0$, $E_0$, and $B_0$ describe the amplitude, center energy, and the broadening of the interband transition, respectively; $W_R$ sets the width of the function on the right side of the center energy; $P_R$ ($A_R$) is the relative position (amplitude) of the control point that separates the two polynomials of the right side ($0 \leq P_R(A_R) \leq 1$, where 0 (1) corresponds to the position of $W_R$ ($A_0$) ); $O_{2R}$ is the coefficient of the 2$^{nd}$ order terms in the polynomial on the right side. In the fitting done on our Ag films, all the left side values are set to zero since a "band gap" is assumed.

### S3: The fitting parameters and fitted results

**Table S1. Fitted values of the variables in Equation 1.**

|  | Epitaxial (38.4 nm) | Epitaxial (44.8 nm) with oxide cap | Thermal (48 nm) |
|---|---|---|---|
| $A_D$ | 2079 (fixed) | 713.08 (fixed) | 38.873 (fixed) |
| $B_D$ | 0.027 ± 0.0009 | 0.0433 ± 0.0037 | 0.2954 ± 0.006 |
| $A_L$ | 1.1397 ± 0.112 | 1.0971 ± 0.142 | 1.317 ± 0.0663 |
| $E_L$ | 6.2297 ± 0.0686 | 6.1925 ± 0.113 | 6.4599 ± 0.125 |
| $B_L$ | 2.5528 ± 0.16 | 2.3805 ± 0.221 | 3.3822 ± 0.113 |
| $A_0$ | 7.7515 ± 0.151 | 8.2683 ± 0.205 | 7.4622 ± 0.118 |
| $E_0$ | 4.0602 ± 0.00228 | 4.0538 ± 0.00752 | 4.0509 ± 0.00315 |
| $B_0$ | 0.16565 ± 0.00228 | 0.16509 ± 0.00262 | 0.14462 ± 0.000349 |
| $W_R$ | 3.6288 ± 0.196 | 4.1139 ± 0.334 | 3.5569 ± 0.0276 |
| $P_R$ | 0.96059 ± 0.0125 | 0.96803 ± 0.0121 | 0.99831 ± 0.00317 |
| $A_R$ | 0.5 (fixed) | 0.5 (fixed) | 0.5 (fixed) |
| $O_{2R}$ | 0 (fixed) | 0 (fixed) | 0 (fixed) |

In Table S1 we provide three examples of the fitting parameters used to extract the optical constants of the 40 nm uncapped epitaxial film, the 50 nm oxide capped epitaxial film, and the 50 nm uncapped thermal film as presented in the main paper (Figure 2). The variables listed in Table S1 are allowed to vary in the regression fitting based on the three-component model (except the ones labeled as "fixed"). Some values are fixed in order to minimize correlations between parameters which can lead to large error bars. In particular, the amplitude, $A_D$, of the Drude component obtained in the first round of regression fits, where it was allowed to vary, was held fixed in the second round. By doing so, we minimized the error bars and correlations between parameters listed in Table S1.

We plot the three components of the fitting function individually in Figure S1. Inspecting this figure allow us to identify the frequency range in which each component makes a dominant contribution. The Drude component is most responsible for loss in energy below 2.3 eV, above which the additional Lorentzian starts to make an appreciable contribution.

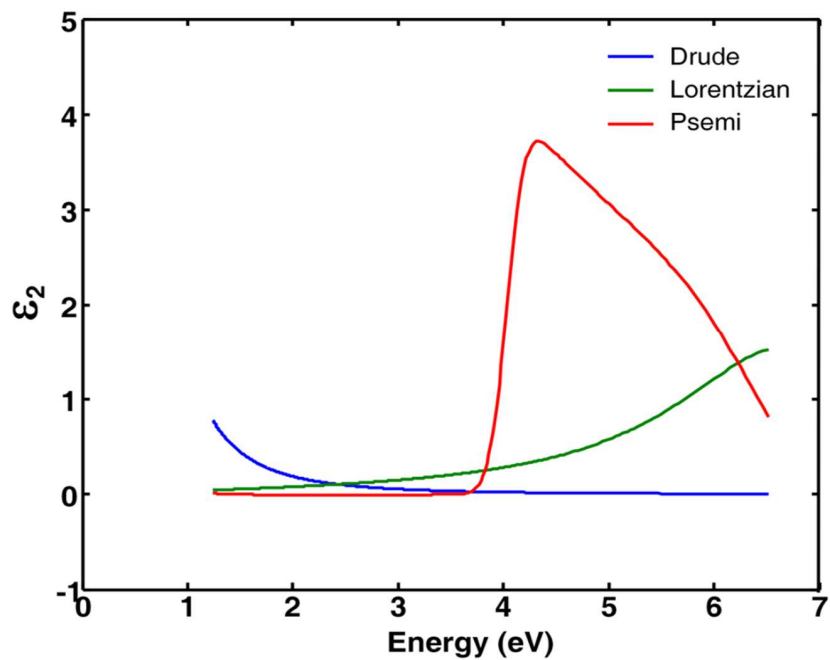

**Figure S1. Components of the parametric model used in the regression fit of the ellipsometry data of silver films.**

In Figure S2, we added the fit residue to the fitted optical constant and compared it with the Johnson and Christy (JC) data with errors included. In the low energy region below the transition threshold, the optical constants extracted from our epitaxial film lie within the large error of the JC data. We would like to emphasize that the parameter values themselves do not necessarily carry any physical meaning. In order to extract useful information on the optical properties of the sample, we need to look at the numerical values of the dielectric constants. We provide a complete list of values of the fitted dielectric constants on the 40 $nm$ uncapped epitaxially grown silver film in Table 2S of section S6.

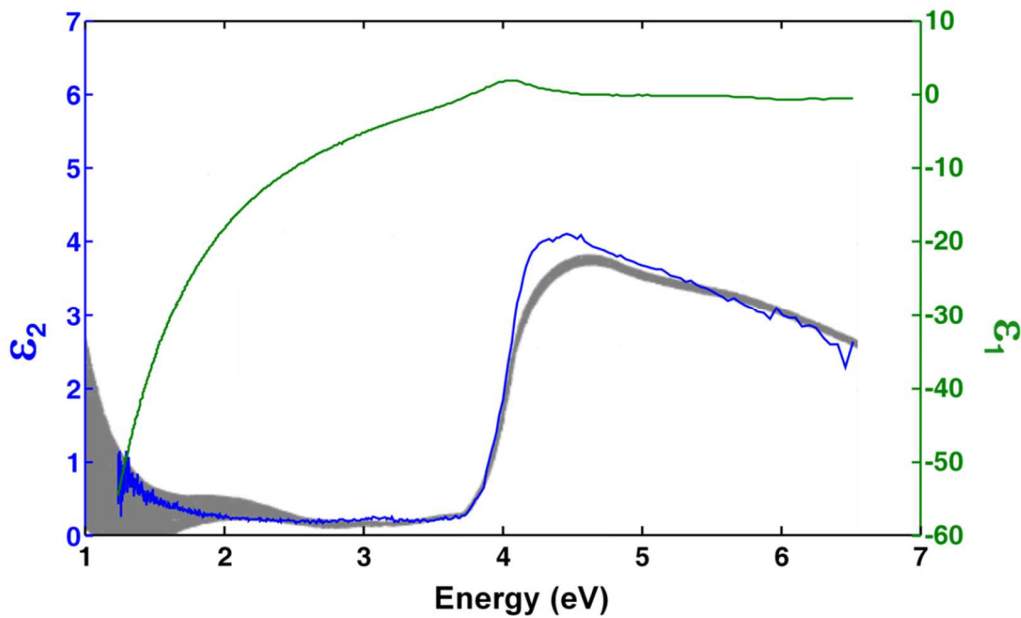

**Figure S2. Optical Constants of the 40nm uncapped epitaxial silver film plotted against the Johnson and Christy data on $\varepsilon_2$ (gray plot) including errors.**

**S4: Multi-sample Analysis**

We investigated a total of six silver films: three epitaxially grown and three thermally deposited. The extracted imaginary part of the optical constants ($\varepsilon_2$) of each film is plotted in Figure S3. We can see that all epitaxially grown films, capped or uncapped, consistently have lower values than the Johnson and Christy (JC) values over a wide energy range in the visible. On the other hand, all the thermally deposited films have higher values than the JC values in the same region. Another reoccurring feature in the fit residue is the absence (presence) of a peak around the energy of 3.7 eV in epitaxial (thermal) films. This peak results from grain boundaries in the thermal films. This set of data further proves the robustness of our chosen model and validates our retrieved optical constants as we obtain consistent fitting results across six different silver films. We note that there are variations in the quality of the epitaxial films. Our AFM images show that the film quality was relatively poor for the one presented in Fig. S3d. The extracted optical constants also yielded higher loss in this particular film.

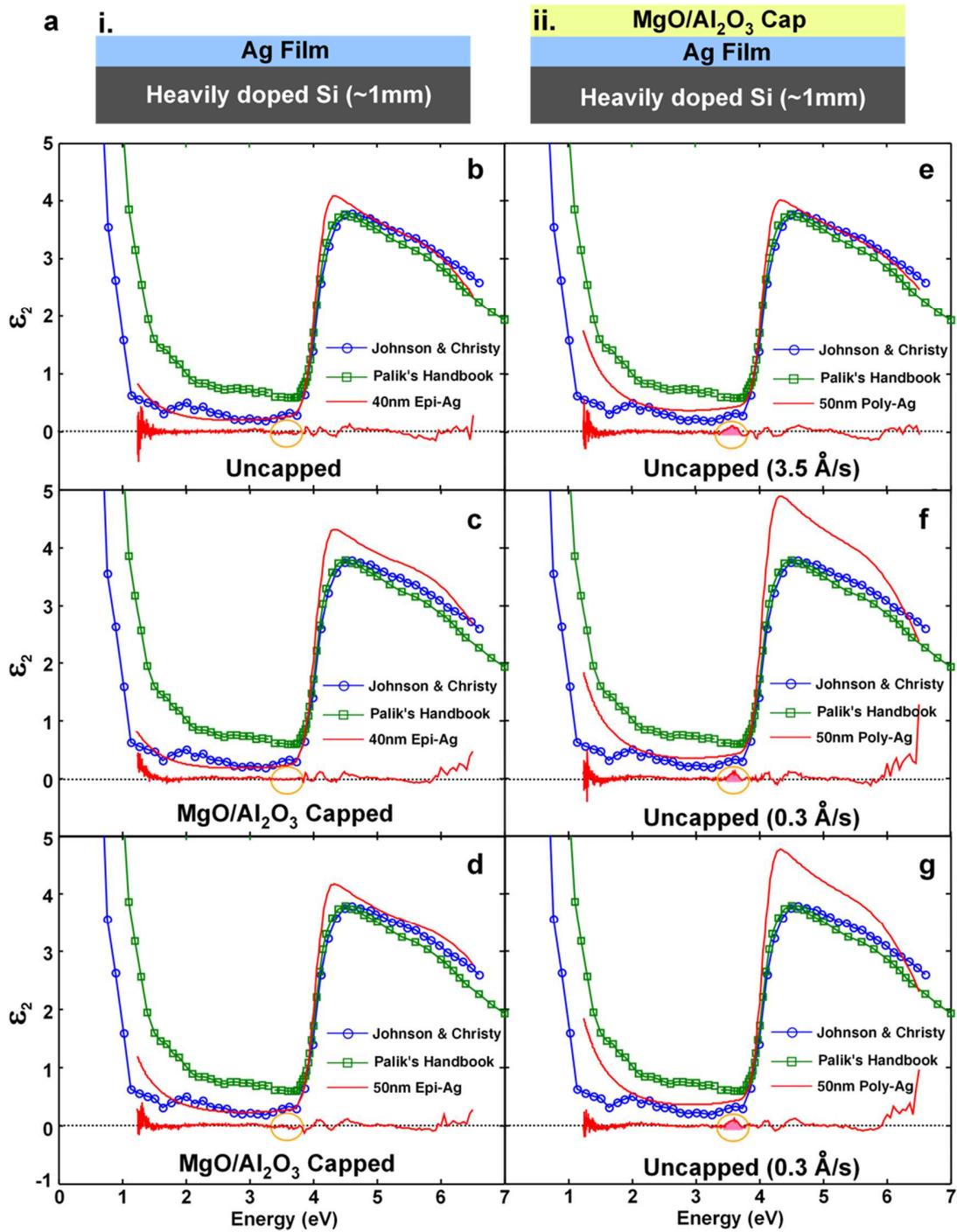

**Figure S3. SE measurements of multiple silver films. a.** Layer structures of uncapped and capped films. **b-d.** Epitaxially grown silver films and **e-g.** Thermally deposited silver films plotted against optical constant data from Johnson and Christy's 1972 paper and Palik's Handbook of optical constants.

**S5: Uniqueness Test**

The mean square error (MSE) is a crucial parameter to evaluate to quantify the quality of our fit. However, a small MSE alone is not a definitive proof that the model used in the regression fitting process is robust. If a model contains highly correlated parameters, then it is possible to have multiple solutions with similarly low MSE values. A strong correlation can exist between thickness and optical constants in absorptive thin films during the regression fitting process and lead to unreliable fitted values. We can eliminate this correlation by using a minimal number of varying parameters in the model and by requiring the fit to comply with the physical Kramers-Kronig (K-K) constraints. To verify that the final fit solution is truly unique, we need to conduct a uniqueness test showing that there is indeed a best fit at a singular value of a chosen parameter. The parameter we chose to perform the uniqueness test on is the thickness of the film. By fixing the thickness of the film at a range of various values, while letting the other parameters vary during the regression fitting process, we calculated the MSE of each final fit result. The MSE is then plotted as a function of thickness. We show in Figure S4 that there is a global minimum within the thickness range for each of the six films investigated. In addition, these minima all occur within 1-2 nm of the film thickness extracted independently from calibration during the growth process. From these uniqueness tests, we conclude that our parameterized model is indeed robust and the retrieved optical constants are valid.

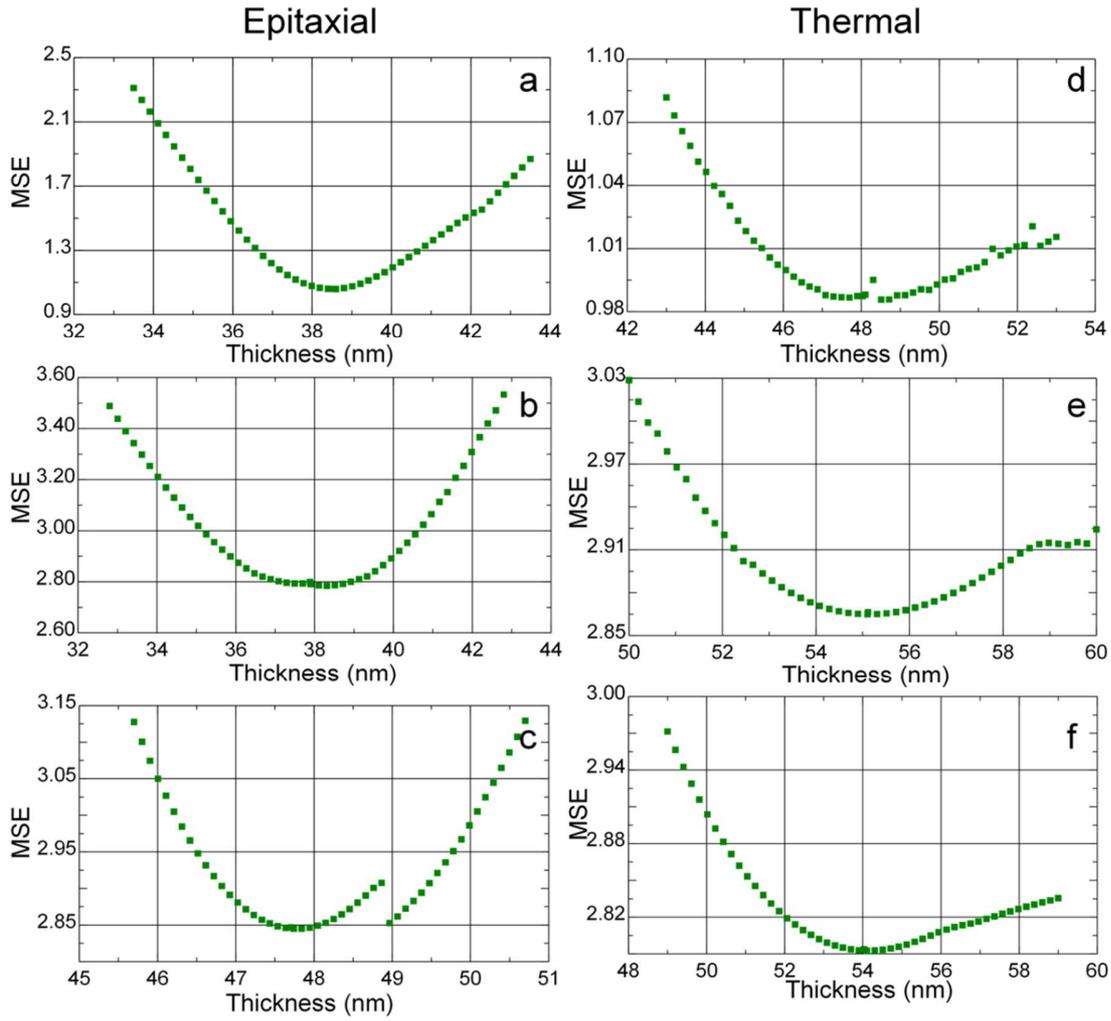

**Figure S4. Uniqueness test of multiple silver films. a-c.** Epitaxially grown silver films corresponding those listed in Figure 3S b-d and **d-f.** Thermally deposited silver films corresponding to those listed in Figure 3S e-g.

## S6: Look-up Table for the Optical Constants of Epitaxially Grown Silver Films

We provide the tabulated optical constants extracted from the 40-nm epitaxial film below. In principle, an analytical form of the fit result can be provided if the exact expression for $\varepsilon_{Psemi}$ is available.

**Table S2. The complete optical constants look-up table for epitaxially grown Ag films**

| Energy(eV) | $\varepsilon_1$ | $\varepsilon_2$ | Energy(eV) | $\varepsilon_1$ | $\varepsilon_2$ | Energy(eV) | $\varepsilon_1$ | $\varepsilon_2$ |
|---|---|---|---|---|---|---|---|---|
| 1.2407 | -54.458 | 0.83494 | 1.3459 | -45.655 | 0.66709 | 1.4714 | -37.515 | 0.52626 |
| 1.2427 | -54.276 | 0.83131 | 1.3482 | -45.485 | 0.66401 | 1.4742 | -37.359 | 0.52371 |
| 1.2446 | -54.094 | 0.82769 | 1.3505 | -45.316 | 0.66094 | 1.4769 | -37.203 | 0.52117 |
| 1.2466 | -53.912 | 0.82409 | 1.3528 | -45.147 | 0.65789 | 1.4797 | -37.048 | 0.51864 |
| 1.2485 | -53.73 | 0.82049 | 1.3552 | -44.979 | 0.65484 | 1.4825 | -36.893 | 0.51612 |
| 1.2505 | -53.549 | 0.8169 | 1.3575 | -44.81 | 0.65181 | 1.4853 | -36.738 | 0.51361 |
| 1.2525 | -53.367 | 0.81333 | 1.3598 | -44.642 | 0.64879 | 1.4881 | -36.583 | 0.51111 |
| 1.2544 | -53.187 | 0.80977 | 1.3622 | -44.475 | 0.64578 | 1.4909 | -36.429 | 0.50862 |
| 1.2564 | -53.006 | 0.80621 | 1.3645 | -44.307 | 0.64277 | 1.4937 | -36.275 | 0.50614 |
| 1.2584 | -52.825 | 0.80267 | 1.3669 | -44.14 | 0.63978 | 1.4966 | -36.121 | 0.50367 |
| 1.2604 | -52.645 | 0.79915 | 1.3692 | -43.973 | 0.6368 | 1.4994 | -35.968 | 0.50122 |
| 1.2624 | -52.465 | 0.79563 | 1.3716 | -43.806 | 0.63383 | 1.5023 | -35.815 | 0.49877 |
| 1.2644 | -52.286 | 0.79212 | 1.374 | -43.64 | 0.63087 | 1.5052 | -35.662 | 0.49633 |
| 1.2664 | -52.107 | 0.78863 | 1.3764 | -43.474 | 0.62793 | 1.5081 | -35.509 | 0.4939 |
| 1.2684 | -51.927 | 0.78514 | 1.3788 | -43.308 | 0.62499 | 1.511 | -35.357 | 0.49148 |
| 1.2705 | -51.749 | 0.78167 | 1.3812 | -43.142 | 0.62206 | 1.5139 | -35.205 | 0.48908 |
| 1.2725 | -51.57 | 0.77821 | 1.3836 | -42.977 | 0.61915 | 1.5168 | -35.053 | 0.48668 |
| 1.2745 | -51.392 | 0.77476 | 1.386 | -42.812 | 0.61624 | 1.5197 | -34.901 | 0.48429 |
| 1.2766 | -51.214 | 0.77132 | 1.3885 | -42.647 | 0.61335 | 1.5227 | -34.75 | 0.48192 |
| 1.2786 | -51.036 | 0.76789 | 1.3909 | -42.482 | 0.61046 | 1.5256 | -34.599 | 0.47955 |
| 1.2807 | -50.858 | 0.76447 | 1.3934 | -42.318 | 0.60759 | 1.5286 | -34.448 | 0.47719 |
| 1.2828 | -50.681 | 0.76107 | 1.3958 | -42.154 | 0.60473 | 1.5316 | -34.298 | 0.47484 |
| 1.2848 | -50.504 | 0.75767 | 1.3983 | -41.99 | 0.60187 | 1.5346 | -34.148 | 0.47251 |
| 1.2869 | -50.327 | 0.75429 | 1.4008 | -41.827 | 0.59903 | 1.5376 | -33.998 | 0.47018 |
| 1.289 | -50.151 | 0.75091 | 1.4033 | -41.664 | 0.5962 | 1.5406 | -33.848 | 0.46786 |
| 1.2911 | -49.975 | 0.74755 | 1.4058 | -41.501 | 0.59338 | 1.5436 | -33.699 | 0.46556 |
| 1.2932 | -49.799 | 0.7442 | 1.4083 | -41.338 | 0.59057 | 1.5467 | -33.55 | 0.46326 |
| 1.2953 | -49.623 | 0.74086 | 1.4108 | -41.176 | 0.58777 | 1.5497 | -33.401 | 0.46097 |
| 1.2974 | -49.448 | 0.73754 | 1.4133 | -41.013 | 0.58498 | 1.5528 | -33.252 | 0.4587 |
| 1.2996 | -49.272 | 0.73422 | 1.4159 | -40.852 | 0.5822 | 1.5558 | -33.104 | 0.45643 |
| 1.3017 | -49.097 | 0.73091 | 1.4184 | -40.69 | 0.57943 | 1.5589 | -32.956 | 0.45417 |
| 1.3038 | -48.923 | 0.72762 | 1.421 | -40.529 | 0.57668 | 1.562 | -32.808 | 0.45193 |
| 1.306 | -48.748 | 0.72433 | 1.4235 | -40.368 | 0.57393 | 1.5652 | -32.661 | 0.44969 |
| 1.3081 | -48.574 | 0.72106 | 1.4261 | -40.207 | 0.57119 | 1.5683 | -32.514 | 0.44746 |
| 1.3103 | -48.4 | 0.7178 | 1.4287 | -40.046 | 0.56847 | 1.5714 | -32.367 | 0.44525 |
| 1.3125 | -48.227 | 0.71455 | 1.4313 | -39.886 | 0.56575 | 1.5746 | -32.22 | 0.44304 |
| 1.3147 | -48.054 | 0.71131 | 1.4339 | -39.726 | 0.56304 | 1.5777 | -32.074 | 0.44084 |
| 1.3168 | -47.88 | 0.70808 | 1.4365 | -39.566 | 0.56035 | 1.5809 | -31.928 | 0.43865 |
| 1.319 | -47.708 | 0.70486 | 1.4391 | -39.407 | 0.55766 | 1.5841 | -31.782 | 0.43648 |
| 1.3212 | -47.535 | 0.70165 | 1.4418 | -39.248 | 0.55499 | 1.5873 | -31.637 | 0.43431 |
| 1.3234 | -47.363 | 0.69845 | 1.4444 | -39.089 | 0.55233 | 1.5905 | -31.492 | 0.43215 |
| 1.3256 | -47.191 | 0.69527 | 1.4471 | -38.93 | 0.54967 | 1.5938 | -31.347 | 0.43 |
| 1.3279 | -47.019 | 0.69209 | 1.4497 | -38.772 | 0.54703 | 1.597 | -31.202 | 0.42787 |
| 1.3301 | -46.848 | 0.68893 | 1.4524 | -38.614 | 0.5444 | 1.6003 | -31.058 | 0.42574 |
| 1.3323 | -46.676 | 0.68578 | 1.4551 | -38.456 | 0.54177 | 1.6035 | -30.913 | 0.42362 |
| 1.3346 | -46.505 | 0.68263 | 1.4578 | -38.298 | 0.53916 | 1.6068 | -30.77 | 0.42151 |
| 1.3368 | -46.335 | 0.6795 | 1.4605 | -38.141 | 0.53656 | 1.6101 | -30.626 | 0.41941 |
| 1.3391 | -46.164 | 0.67638 | 1.4632 | -37.984 | 0.53397 | 1.6134 | -30.483 | 0.41733 |
| 1.3414 | -45.994 | 0.67327 | 1.4659 | -37.827 | 0.53139 | 1.6168 | -30.34 | 0.41525 |
| 1.3436 | -45.824 | 0.67018 | 1.4687 | -37.671 | 0.52882 | 1.6201 | -30.197 | 0.41318 |

| Energy(eV) | $\varepsilon_1$ | $\varepsilon_2$ | Energy(eV) | $\varepsilon_1$ | $\varepsilon_2$ | Energy(eV) | $\varepsilon_1$ | $\varepsilon_2$ |
|---|---|---|---|---|---|---|---|---|
| 1.6235 | -30.055 | 0.41112 | 1.8114 | -23.285 | 0.32048 | 2.0495 | -17.211 | 0.2536 |
| 1.6268 | -29.912 | 0.40907 | 1.8157 | -23.157 | 0.31891 | 2.0549 | -17.097 | 0.25251 |
| 1.6302 | -29.77 | 0.40703 | 1.8199 | -23.029 | 0.31735 | 2.0603 | -16.982 | 0.25143 |
| 1.6336 | -29.629 | 0.405 | 1.8241 | -22.901 | 0.3158 | 2.0658 | -16.869 | 0.25035 |
| 1.637 | -29.487 | 0.40298 | 1.8284 | -22.774 | 0.31426 | 2.0713 | -16.755 | 0.24928 |
| 1.6405 | -29.346 | 0.40097 | 1.8327 | -22.647 | 0.31272 | 2.0768 | -16.642 | 0.24823 |
| 1.6439 | -29.206 | 0.39897 | 1.837 | -22.52 | 0.3112 | 2.0824 | -16.529 | 0.24718 |
| 1.6474 | -29.065 | 0.39698 | 1.8413 | -22.393 | 0.30969 | 2.0879 | -16.416 | 0.24614 |
| 1.6508 | -28.925 | 0.395 | 1.8457 | -22.267 | 0.30819 | 2.0936 | -16.303 | 0.24512 |
| 1.6543 | -28.785 | 0.39302 | 1.8501 | -22.141 | 0.30669 | 2.0992 | -16.191 | 0.2441 |
| 1.6578 | -28.645 | 0.39106 | 1.8545 | -22.015 | 0.30521 | 2.1049 | -16.079 | 0.24309 |
| 1.6613 | -28.506 | 0.38911 | 1.8589 | -21.889 | 0.30373 | 2.1106 | -15.967 | 0.2421 |
| 1.6649 | -28.367 | 0.38717 | 1.8633 | -21.764 | 0.30227 | 2.1163 | -15.856 | 0.24111 |
| 1.6684 | -28.228 | 0.38524 | 1.8678 | -21.639 | 0.30081 | 2.1221 | -15.745 | 0.24013 |
| 1.672 | -28.089 | 0.38331 | 1.8723 | -21.515 | 0.29937 | 2.1279 | -15.634 | 0.23916 |
| 1.6756 | -27.951 | 0.3814 | 1.8768 | -21.39 | 0.29793 | 2.1337 | -15.523 | 0.2382 |
| 1.6792 | -27.813 | 0.3795 | 1.8813 | -21.266 | 0.2965 | 2.1396 | -15.413 | 0.23725 |
| 1.6828 | -27.675 | 0.3776 | 1.8858 | -21.142 | 0.29508 | 2.1455 | -15.303 | 0.23631 |
| 1.6864 | -27.538 | 0.37572 | 1.8904 | -21.019 | 0.29368 | 2.1514 | -15.193 | 0.23539 |
| 1.69 | -27.4 | 0.37384 | 1.895 | -20.895 | 0.29228 | 2.1574 | -15.084 | 0.23447 |
| 1.6937 | -27.263 | 0.37198 | 1.8996 | -20.772 | 0.29089 | 2.1634 | -14.974 | 0.23356 |
| 1.6974 | -27.127 | 0.37012 | 1.9043 | -20.65 | 0.28951 | 2.1694 | -14.865 | 0.23266 |
| 1.7011 | -26.99 | 0.36828 | 1.9089 | -20.527 | 0.28814 | 2.1755 | -14.757 | 0.23177 |
| 1.7048 | -26.854 | 0.36644 | 1.9136 | -20.405 | 0.28677 | 2.1816 | -14.648 | 0.23089 |
| 1.7085 | -26.719 | 0.36461 | 1.9183 | -20.283 | 0.28542 | 2.1877 | -14.54 | 0.23002 |
| 1.7122 | -26.583 | 0.3628 | 1.923 | -20.161 | 0.28408 | 2.1939 | -14.432 | 0.22916 |
| 1.716 | -26.448 | 0.36099 | 1.9278 | -20.04 | 0.28275 | 2.2001 | -14.325 | 0.22832 |
| 1.7198 | -26.313 | 0.35919 | 1.9326 | -19.919 | 0.28142 | 2.2063 | -14.217 | 0.22748 |
| 1.7236 | -26.178 | 0.3574 | 1.9374 | -19.798 | 0.28011 | 2.2126 | -14.11 | 0.22665 |
| 1.7274 | -26.044 | 0.35563 | 1.9422 | -19.678 | 0.2788 | 2.2189 | -14.003 | 0.22583 |
| 1.7312 | -25.91 | 0.35386 | 1.9471 | -19.558 | 0.27751 | 2.2252 | -13.897 | 0.22502 |
| 1.735 | -25.776 | 0.3521 | 1.9519 | -19.438 | 0.27622 | 2.2316 | -13.791 | 0.22422 |
| 1.7389 | -25.642 | 0.35035 | 1.9568 | -19.318 | 0.27495 | 2.238 | -13.685 | 0.22344 |
| 1.7428 | -25.509 | 0.34861 | 1.9618 | -19.198 | 0.27368 | 2.2445 | -13.579 | 0.22266 |
| 1.7467 | -25.376 | 0.34688 | 1.9667 | -19.079 | 0.27242 | 2.251 | -13.473 | 0.22189 |
| 1.7506 | -25.243 | 0.34515 | 1.9717 | -18.961 | 0.27118 | 2.2575 | -13.368 | 0.22114 |
| 1.7545 | -25.111 | 0.34344 | 1.9767 | -18.842 | 0.26994 | 2.2641 | -13.263 | 0.22039 |
| 1.7584 | -24.978 | 0.34174 | 1.9817 | -18.724 | 0.26871 | 2.2707 | -13.158 | 0.21966 |
| 1.7624 | -24.847 | 0.34005 | 1.9868 | -18.606 | 0.26749 | 2.2774 | -13.054 | 0.21893 |
| 1.7664 | -24.715 | 0.33836 | 1.9918 | -18.488 | 0.26628 | 2.2841 | -12.95 | 0.21822 |
| 1.7704 | -24.584 | 0.33669 | 1.9969 | -18.37 | 0.26508 | 2.2908 | -12.846 | 0.21751 |
| 1.7744 | -24.452 | 0.33503 | 2.0021 | -18.253 | 0.26389 | 2.2975 | -12.742 | 0.21682 |
| 1.7785 | -24.322 | 0.33337 | 2.0072 | -18.136 | 0.26271 | 2.3044 | -12.639 | 0.21614 |
| 1.7825 | -24.191 | 0.33173 | 2.0124 | -18.02 | 0.26154 | 2.3112 | -12.536 | 0.21546 |
| 1.7866 | -24.061 | 0.33009 | 2.0176 | -17.903 | 0.26038 | 2.3181 | -12.433 | 0.2148 |
| 1.7907 | -23.931 | 0.32846 | 2.0229 | -17.787 | 0.25922 | 2.325 | -12.33 | 0.21415 |
| 1.7948 | -23.801 | 0.32685 | 2.0281 | -17.671 | 0.25808 | 2.332 | -12.228 | 0.21351 |
| 1.7989 | -23.672 | 0.32524 | 2.0334 | -17.556 | 0.25695 | 2.339 | -12.126 | 0.21289 |
| 1.8031 | -23.543 | 0.32364 | 2.0387 | -17.441 | 0.25582 | 2.3461 | -12.024 | 0.21227 |
| 1.8073 | -23.414 | 0.32205 | 2.0441 | -17.326 | 0.25471 | 2.3532 | -11.923 | 0.21166 |

| Energy(eV) | $\varepsilon_1$ | $\varepsilon_2$ | Energy(eV) | $\varepsilon_1$ | $\varepsilon_2$ | Energy(eV) | $\varepsilon_1$ | $\varepsilon_2$ |
|---|---|---|---|---|---|---|---|---|
| 2.3603 | -11.821 | 0.21107 | 2.783 | -7.0649 | 0.19718 | 3.3904 | -2.6758 | 0.23001 |
| 2.3675 | -11.72 | 0.21048 | 2.793 | -6.9753 | 0.19728 | 3.4053 | -2.5842 | 0.23148 |
| 2.3747 | -11.619 | 0.20991 | 2.8031 | -6.8859 | 0.19739 | 3.4203 | -2.4919 | 0.23301 |
| 2.382 | -11.519 | 0.20935 | 2.8132 | -6.7966 | 0.19751 | 3.4354 | -2.3988 | 0.23458 |
| 2.3893 | -11.419 | 0.2088 | 2.8235 | -6.7076 | 0.19766 | 3.4507 | -2.3048 | 0.23621 |
| 2.3967 | -11.319 | 0.20826 | 2.8338 | -6.6186 | 0.19782 | 3.4661 | -2.21 | 0.2379 |
| 2.4041 | -11.219 | 0.20773 | 2.8441 | -6.5299 | 0.198 | 3.4816 | -2.1141 | 0.23966 |
| 2.4116 | -11.119 | 0.20721 | 2.8546 | -6.4413 | 0.1982 | 3.4973 | -2.017 | 0.24149 |
| 2.4191 | -11.02 | 0.20671 | 2.8651 | -6.3529 | 0.19842 | 3.5131 | -1.9186 | 0.24342 |
| 2.4266 | -10.921 | 0.20621 | 2.8757 | -6.2646 | 0.19866 | 3.5291 | -1.8188 | 0.24546 |
| 2.4342 | -10.822 | 0.20573 | 2.8864 | -6.1765 | 0.19892 | 3.5452 | -1.7173 | 0.24765 |
| 2.4419 | -10.724 | 0.20526 | 2.8972 | -6.0885 | 0.19919 | 3.5614 | -1.6139 | 0.25004 |
| 2.4496 | -10.626 | 0.20481 | 2.908 | -6.0007 | 0.19949 | 3.5778 | -1.5084 | 0.25269 |
| 2.4573 | -10.528 | 0.20436 | 2.919 | -5.913 | 0.19981 | 3.5944 | -1.4005 | 0.25571 |
| 2.4651 | -10.43 | 0.20393 | 2.93 | -5.8254 | 0.20015 | 3.6111 | -1.2899 | 0.25922 |
| 2.473 | -10.332 | 0.2035 | 2.9411 | -5.7379 | 0.20051 | 3.628 | -1.1761 | 0.26342 |
| 2.4808 | -10.235 | 0.20309 | 2.9522 | -5.6506 | 0.20089 | 3.645 | -1.0589 | 0.26858 |
| 2.4888 | -10.138 | 0.2027 | 2.9635 | -5.5633 | 0.20129 | 3.6622 | -0.93759 | 0.27504 |
| 2.4968 | -10.041 | 0.20231 | 2.9749 | -5.4762 | 0.20171 | 3.6795 | -0.81184 | 0.28329 |
| 2.5048 | -9.9448 | 0.20194 | 2.9863 | -5.3892 | 0.20216 | 3.697 | -0.68109 | 0.29393 |
| 2.5129 | -9.8485 | 0.20158 | 2.9978 | -5.3022 | 0.20263 | 3.7147 | -0.54485 | 0.30775 |
| 2.5211 | -9.7525 | 0.20123 | 3.0094 | -5.2154 | 0.20313 | 3.7326 | -0.40265 | 0.32574 |
| 2.5293 | -9.6567 | 0.2009 | 3.0211 | -5.1286 | 0.20364 | 3.7506 | -0.25413 | 0.34912 |
| 2.5376 | -9.5611 | 0.20057 | 3.0329 | -5.0419 | 0.20418 | 3.7688 | -0.09907 | 0.37933 |
| 2.5459 | -9.4658 | 0.20027 | 3.0448 | -4.9552 | 0.20475 | 3.7871 | 0.062485 | 0.41808 |
| 2.5542 | -9.3707 | 0.19997 | 3.0568 | -4.8686 | 0.20534 | 3.8057 | 0.2302 | 0.46725 |
| 2.5627 | -9.2758 | 0.19969 | 3.0689 | -4.782 | 0.20596 | 3.8244 | 0.4033 | 0.52892 |
| 2.5711 | -9.1812 | 0.19942 | 3.0811 | -4.6954 | 0.2066 | 3.8433 | 0.5805 | 0.60523 |
| 2.5797 | -9.0868 | 0.19916 | 3.0933 | -4.6089 | 0.20727 | 3.8624 | 0.75994 | 0.69829 |
| 2.5883 | -8.9926 | 0.19892 | 3.1057 | -4.5224 | 0.20797 | 3.8817 | 0.93912 | 0.81001 |
| 2.5969 | -8.8986 | 0.19869 | 3.1182 | -4.4359 | 0.20869 | 3.9012 | 1.1149 | 0.94189 |
| 2.6056 | -8.8049 | 0.19847 | 3.1308 | -4.3493 | 0.20944 | 3.9209 | 1.2835 | 1.0948 |
| 2.6144 | -8.7114 | 0.19827 | 3.1434 | -4.2627 | 0.21022 | 3.9408 | 1.4407 | 1.269 |
| 2.6232 | -8.6182 | 0.19808 | 3.1562 | -4.1761 | 0.21103 | 3.9609 | 1.582 | 1.4634 |
| 2.6321 | -8.5251 | 0.19791 | 3.1691 | -4.0894 | 0.21187 | 3.9812 | 1.703 | 1.6762 |
| 2.6411 | -8.4323 | 0.19775 | 3.1821 | -4.0026 | 0.21274 | 4.0017 | 1.7995 | 1.904 |
| 2.6501 | -8.3397 | 0.19761 | 3.1952 | -3.9158 | 0.21365 | 4.0224 | 1.8679 | 2.1428 |
| 2.6591 | -8.2473 | 0.19748 | 3.2084 | -3.8288 | 0.21458 | 4.0433 | 1.9057 | 2.3871 |
| 2.6683 | -8.1551 | 0.19736 | 3.2217 | -3.7417 | 0.21555 | 4.0645 | 1.9118 | 2.6312 |
| 2.6775 | -8.0631 | 0.19726 | 3.2351 | -3.6544 | 0.21655 | 4.0859 | 1.8863 | 2.8689 |
| 2.6867 | -7.9714 | 0.19717 | 3.2486 | -3.567 | 0.21758 | 4.1074 | 1.831 | 3.0941 |
| 2.6961 | -7.8798 | 0.1971 | 3.2623 | -3.4793 | 0.21865 | 4.1293 | 1.7491 | 3.3016 |
| 2.7054 | -7.7885 | 0.19705 | 3.276 | -3.3914 | 0.21976 | 4.1513 | 1.6451 | 3.4869 |
| 2.7149 | -7.6974 | 0.19701 | 3.2899 | -3.3032 | 0.2209 | 4.1736 | 1.5243 | 3.6469 |
| 2.7244 | -7.6064 | 0.19699 | 3.3039 | -3.2148 | 0.22208 | 4.1962 | 1.3924 | 3.7801 |
| 2.734 | -7.5157 | 0.19698 | 3.318 | -3.126 | 0.2233 | 4.2189 | 1.255 | 3.8865 |
| 2.7437 | -7.4251 | 0.19699 | 3.3322 | -3.0369 | 0.22456 | 4.242 | 1.1174 | 3.9672 |
| 2.7534 | -7.3348 | 0.19701 | 3.3466 | -2.9474 | 0.22586 | 4.2652 | 0.98395 | 4.0247 |
| 2.7632 | -7.2446 | 0.19705 | 3.3611 | -2.8574 | 0.2272 | 4.2888 | 0.858 | 4.0621 |
| 2.7731 | -7.1547 | 0.19711 | 3.3757 | -2.7669 | 0.22858 | 4.3126 | 0.74184 | 4.0829 |

| Energy(eV) | $\varepsilon_1$ | $\varepsilon_2$ | Energy(eV) | $\varepsilon_1$ | $\varepsilon_2$ | Energy(eV) | $\varepsilon_1$ | $\varepsilon_2$ |
|---|---|---|---|---|---|---|---|---|
| 4.3366 | 0.63668 | 4.0905 | 5.8324 | -0.3998 | 3.1751 | | | |
| 4.361 | 0.54283 | 4.0884 | 5.8764 | -0.41951 | 3.1403 | | | |
| 4.3856 | 0.4599 | 4.0792 | 5.9211 | -0.43989 | 3.1026 | | | |
| 4.4104 | 0.38705 | 4.0653 | 5.9665 | -0.46074 | 3.0618 | | | |
| 4.4356 | 0.32317 | 4.0485 | 6.254 | -0.57687 | 2.7367 | | | |
| 4.4611 | 0.26707 | 4.0298 | 6.3046 | -0.59067 | 2.6677 | | | |
| 4.4868 | 0.21762 | 4.0102 | 6.3561 | -0.60155 | 2.5944 | | | |
| 4.5129 | 0.17381 | 3.9901 | 6.4084 | -0.60896 | 2.5169 | | | |
| 4.5392 | 0.13478 | 3.9698 | 6.4616 | -0.61238 | 2.4354 | | | |
| 4.5659 | 0.099826 | 3.9496 | 6.5156 | -0.6113 | 2.35 | | | |
| 4.5928 | 0.068379 | 3.9293 | | | | | | |
| 4.6201 | 0.039984 | 3.9091 | | | | | | |
| 4.6477 | 0.014276 | 3.8889 | | | | | | |
| 4.6757 | -0.00905 | 3.8688 | | | | | | |
| 4.704 | -0.03024 | 3.8487 | | | | | | |
| 4.7326 | -0.04952 | 3.8287 | | | | | | |
| 4.7616 | -0.06708 | 3.8088 | | | | | | |
| 4.7909 | -0.08309 | 3.7889 | | | | | | |
| 4.8206 | -0.09770 | 3.7691 | | | | | | |
| 4.8507 | -0.11106 | 3.7494 | | | | | | |
| 4.8811 | -0.12329 | 3.7297 | | | | | | |
| 4.9119 | -0.13453 | 3.7102 | | | | | | |
| 4.9432 | -0.14489 | 3.6908 | | | | | | |
| 4.9748 | -0.15447 | 3.6714 | | | | | | |
| 5.0068 | -0.1634 | 3.6522 | | | | | | |
| 5.0392 | -0.17178 | 3.633 | | | | | | |
| 5.0721 | -0.17971 | 3.614 | | | | | | |
| 5.1054 | -0.18729 | 3.595 | | | | | | |
| 5.1391 | -0.19464 | 3.5761 | | | | | | |
| 5.1733 | -0.20185 | 3.5573 | | | | | | |
| 5.2079 | -0.20903 | 3.5385 | | | | | | |
| 5.243 | -0.21628 | 3.5197 | | | | | | |
| 5.2786 | -0.22371 | 3.5008 | | | | | | |
| 5.3146 | -0.23142 | 3.4818 | | | | | | |
| 5.3512 | -0.23953 | 3.4627 | | | | | | |
| 5.3882 | -0.24814 | 3.4434 | | | | | | |
| 5.4258 | -0.25736 | 3.4237 | | | | | | |
| 5.4639 | -0.26728 | 3.4035 | | | | | | |
| 5.5025 | -0.278 | 3.3828 | | | | | | |
| 5.5417 | -0.2896 | 3.3614 | | | | | | |
| 5.5815 | -0.30217 | 3.3391 | | | | | | |
| 5.6218 | -0.31577 | 3.3158 | | | | | | |
| 5.6627 | -0.33044 | 3.2912 | | | | | | |
| 5.7042 | -0.3462 | 3.265 | | | | | | |
| 5.7463 | -0.36305 | 3.2372 | | | | | | |
| 5.789 | -0.38094 | 3.2073 | | | | | | |

**S7: Validating the optical constants of epitaxial *Ag* reported**

The validity of these newly retrieved optical constants is justified in several ways. First, because of the high quality of the epitaxial films, we expect absorption to be lower in the Drude region. We consistently extract lower $\varepsilon_2$ in the 1.8 – 2.5 eV range across measurements performed on three different epitaxial films. In contrast, optical constants extracted from thermally evaporated films using the same fitting procedure show higher loss in the Drude region (see Supplementary Information S4 for the complete data set). Second, we employed a physical model for the optical constants, which ensures that the extracted values comply with the K-K relations and causality. Following this model, we conducted a uniqueness test which confirmed the robustness of the fitting procedure (See Supplementary Information S5). Third, the extracted optical constants are nearly identical for multiple epitaxial films, despite variations in the layered structure. Similar consistency was found for thermal films prepared using different deposition rates, especially in the energy range below the *d*-band transition. Finally, all measurements produce qualitatively and even quantitatively similar residues as shown in Figure 2 in the main text, which are associated with spectral features not captured by our simple analytical model. The residues are significant above the *d*-band transition, where we expect our simple model to break down and be insufficient to describe the complex interband transitions. As discussed in the main text, we speculate that the residues in the energy range immediately below the onset of the *d*-band transition arise from the effect of grain boundaries.